\documentclass[aps,pra,onecolumn,preprint,preprintnumbers,groupaddress,amsmath,amssymb,usenatbib,
nofootinbib,showkeys,showpacs,altaffilletter,11pt]{revtex4}
\usepackage{graphicx}% Include figure files
\usepackage{dcolumn}% Align table columns on decimal point
\usepackage{bm}% bold math
\usepackage{latexsym}
\usepackage{mathrsfs}
\usepackage{subfigure}
\usepackage[colorlinks]{hyperref}

\begin{document}

\title{Interacting closed string tachyon with modified Chaplygin gas and its stability}

\author{Ali R. Amani}
\email{a.r.amani@iauamol.ac.ir }
\affiliation{\centerline{Faculty of Basic Sciences, Department of Physics, Ayatollah Amoli Branch, Islamic Azad University,}\\ P.O. Box 678, Amol, Iran.}

\author{Celia Escamilla-Rivera}
\email{celia\_escamilla@ehu.es}
\affiliation{Fisika Teorikoaren eta Zientziaren Historia Saila, Zientzia
eta Teknologia Fakultatea, Euskal Herriko Unibertsitatea, 644 Posta
Kutxatila, 48080, Bilbao, Spain.}

\author{H. R. Faghani}
\email{h.r.faghani@hotmail.com}
\affiliation{\centerline{Department of Physics, Central Tehran Branch, Islamic Azad University, Tehran, Iran.}}

\date{\today}

\pacs{04.62.+v; 11.10.Ef; 98.80.-k}
\keywords{Closed string tachyon; Chaplygin gas; Non-flat space; Equation of State parameter; Stability.}

%##################################################################################

\begin{abstract}
In this paper, we have considered closed string tachyon model with a constant dilaton field and interacted it
with Chaplygin gas for evaluating cosmology parameters. The model has been studied in $26$-dimensional
that its $22$-dimensional is related to compactification on an internal non-flat space and its other $4$-dimensions
is related to FLRW metric. By taking the internal curvature as
a negative constant, we obtained the closed string tachyon potential as a quartic equation. The tachyon field
and the scale factor have been achieved as functional of time evolution and geometry of curved space where
the behaviour of the scale factor describes an accelerated expansion of the universe. Next, we discussed the stability of our model by introducing a sound speed factor, which one must be, in our case, a positive function.
By drawing sound speed against time evolution we investigated stability conditions for non-flat
universe in its three stages: early, late and future time. As a result we shall see that in these cases remains
an instability at early time and a stability point at late time.
\end{abstract}
\maketitle

%##################################################################################

\section{Introduction}\label{s1}

It is known from so many years ago that the String Theory looks like a highly, good candidate
to describe the physical world. At low-energies it clearly gives rise to General Relativity,
scalar fields and gauge models. In other words, this theory contains in a generic way all the
ingredients that overlay our universe. However, at this energy level exist solutions to the effective
action related with instabilities, so-called tachyons. Of course, the main reason for discussing this
solutions is that they all carry directly over to the superstring theories, where the most well known scenario
concerns the presence of a tachyonic mode on the open string spectrum between pairs of D-branes and
anti-D-branes \cite{ASEN, NLAM}. In order to be in the minimal state of energy, the tachyon rolls down to the minimum of the
potential, and the perturbative approach of the theory becomes reliable. This process is called tachyon
condensation. Notwithstanding, tachyonic modes are not the only solutions on this matter, in fact, in the
bosonic string spectrum exist also the closed string tachyon modes.  The fact that this scenarios
 turns out to sit at an unstable point is unfortunate, but the positive thing is that we can
think in a good minimum elsewhere for the tachyon potential. To get there we know that an expansion of the
tachyon potential around $T=0$ looks like a polynomial and the physics behind can be reliable. \\

All these results leads a wide possibility to construct cosmological models that can reproduce the actual
behaviour of the observable universe in which an accelerated expansion is present. Interesting solutions
emerged as a result of the following studies in the last few years, for example as in \cite{Yang:2005rx, swanson, Adams:2001sv, JMES,
Yang:2005rw} where the closed string tachyon field drives the collapse of the
universe. However, in this attempts an expansion stage is still absent. To defray this line, in
\cite{EscamillaRivera:2011di} the authors considered with the above ideas a compactification of a
critical bosonic string theory with a tachyonic potential into a 4-dimensional flat space-time finding
certain conditions in where with an arbitrary closed string tachyon potential the universe reaches a
maximum size and then undergoes to a stage where collapses as the tachyon arrives to the minimum
of its potential.\\

Regarding to the accelerate stage of the universe, a wide range of research explored the possibility
of introduce certain exotic matter with negative pressure. This acceleration as we know can be consequence
of the dark energy influence, which in some models the ideal candidate to represent it is the extended
Chaplygin gas \cite{AKAM, MCBE, MCBE1, HBBE, LPCH}, a fluid with negative pressure that begins to
dominate the matter content and, at the end, the process of structure formation is driven by cold dark matter
without affecting the previous history of the universe.  This kind of Chaplygin gas cosmology has an interesting
connection to String Theory via the Nambu-Goto action for a D-brane moving in a $(D+2)$-dimensional space-time,
feature than can be regarded to the tachyonic panorama. \\

Whilst fully realistic models are complicate and have yet to be constructed, this is why the simplicity of the
tachyon model coupled with a Chaplygin gas suggested that it may still find some use as a model of accelerated
universe. This slope is currently under intense scrutiny and in here we present an attempt to going further.\\

In our present work we want to choose the closure relation between the above ideas by focusing on the particularity
of the model to unify the closed string tachyon and the Chaplygin gas. According to this point of view,
the model seems to be an interesting lead since the acceleration stage is preserved.
For details of the effects of relaxing these assumptions, we refer the lector to the literature cited below.
In Section \ref{s2} we explain briefly the system of equations that represent the closed string tachyon scenario
considering a gravitational field in a 4-dimensional non-flat Friedmann--Lema\^{\i}tre–-Robertson–-Walker (FLRW) metric. We then present the cosmological equations behind this scenario. In Section \ref{s3} we present the description of the full model in the Hamiltonian formalism.
We show that, despite the additional freedom in this model, is possible to reconstructed the closed string tachyon potential
as we present in Section \ref{s4}. In Section \ref{s5} we adopt the case when we coupled a Chaplygin gas to the closed string tachyon model \cite{EscamillaRivera:2011di}. An interesting cosmological analysis was made in Section \ref{s6}. We conclude in
Section \ref{s7} that this modified model is capable to describe the current acceleration of the present epoch.

%##################################################################################

\section{Closed string tachyon background}\label{s2}

Let us start by considering critical bosonic string theory with a constant dilaton. The
corresponding action is written for the closed
string tachyon field $T$ in $26$-dimensional space-time as the following form  \cite{Yang:2005rw, Yang:2005rx}
\begin{equation}\label{e1}
S=\frac{1}{2\kappa_0^2}\int~d^{26}x\sqrt{-g_{26}}~e^{-2\Phi_0}\left[R-(\nabla T)^2
-2V(T)\right],
\end{equation}
where $\Phi_0$, $T(t)$ and $V(T)$ are the constant dilaton, a rolling tachyon field and the closed string tachyon
potential, respectively. The 26-dimensional metric, $g_{26}$ is in the form,
\begin{equation}\label{e2}
ds^2=g_{\mu\nu}dx^\mu dx^\nu+h_{mn}dy^mdy^n,
\end{equation}
where the first r.h.s term denotes a spatially $3+1$ dimensional FLRW metric and indices $m$, $n$ running from \(4\) to \(25\). Thus non-flat FLRW background is given by,
\begin{eqnarray}\label{e2-1}
ds^2&=&-dt^2+a(t)^2\left(\frac{dr^2}{1-kr^2}+r^2d\Omega^2\right),
\end{eqnarray}
where $k$ denotes the curvature of space i.e., $k=-1,0,1$ are open, flat and closed universe, respectively. Therefore we can write the effective four-dimensional action by compactification on a non-flat internal $22\,d$ space $X_{22}$ as
\begin{eqnarray}\label{e3}
S&=&\frac{1}{2\kappa_0^2}\int d^4x \sqrt{-g} ~Vol(X_{22})e^{-2\Phi_0}\left[R_4+\mathcal{R}-(\nabla T)^2-2V(T)\right],
\end{eqnarray}
where $Vol(X_{22})$ is the volume of $X_{22}$, $\kappa_0^2=8\pi G_{26}$ is the gravitational strength in $26$ dimensions and $\mathcal{R}$ is constant curvature.\\
In this model we will consider a constant volume for the internal compact
space. Thus the effective $4\,d$ action is written as,
\begin{equation}\label{e4}
S=\frac{m_p^2}{2}\int d^4x~\sqrt{-g^E}\left[R_4\left(g^E\right)-(\nabla T)^2-2 \widetilde{V}(T)\right],
\end{equation}
where $m_p^2$ and $\widetilde{V}$ are the reduced $4\,d$ mass of Planck and the effective scalar potential respectively, and they are given as the following form,
\begin{equation}\label{e5}
m_p^2=e^{-2\Phi_0} \frac{Vol(X_{22}) }{\kappa_0^2},
\end{equation}
\begin{equation}\label{e6}
\widetilde{V}(T)= V(T) -\frac{1}{2}\mathcal{R}.
\end{equation}
Now by taking $a(t)=e^{\alpha(t)}$ and $m_p = 1$ we can obtain Friedmann equations for action \eqref{e4} as the following form,
\begin{eqnarray}\label{e7}
\rho_{tot}&=&3 \left( H^2 + \frac{k}{a^2}\right)=3  \left( \dot{\alpha}^2 + \frac{k}{e^{2 \alpha}}\right),\\
p_{tot}&=&- \left(2 \dot{H}+3 H^2 + \frac{k}{a^2}\right)=- \left(2 \ddot{\alpha}+3 \dot{\alpha}^2 + \frac{k}{e^{2 \alpha}}\right),
\end{eqnarray}
and the equation of the closed string tachyon field is,
\begin{equation}\label{e8}
\ddot{T}+3 \dot{\alpha} \dot{T}+ \frac{d\widetilde{V}}{dT}=0.
\end{equation}

%#########################################################################################

\section{Canonical Hamiltonian Background}\label{s3}

In this section we are going to consider the canonical Hamiltonian analysis. For this case, we have to rewrite the lagrangian density of action \eqref{e4} with respect to generalized parameters $\dot{\alpha}$ and $\dot{T}$. Then, the lagrangian density becomes,
\begin{equation}\label{e21}
S(\alpha, T, t) = \int {\mathcal{L} \, dt}
\end{equation}
where
\begin{eqnarray}\label{e22}
\mathcal{L}=\left(-3\dot{\alpha}^2+\frac{1}{2}\dot{T}^2-\widetilde{V}+3 \frac{k}{e^{2\alpha}}\right)e^{3 \alpha},
\end{eqnarray}
and
\begin{eqnarray}\label{e22-1}
\mathcal{L}=\frac{dS(\alpha, T, t)}{dt}=\frac{\partial S}{\partial \alpha} \dot{\alpha}+\frac{\partial S}{\partial T} \dot{T}+\frac{\partial S}{\partial t}.
\end{eqnarray}
By using Hamilton's principle function, we can write canonical Hamiltonian equations and Hamiltonian--Jacobi equation in the following form respectively,
\begin{subequations}
\begin{eqnarray}
\pi_{\alpha}=\frac{\partial S}{\partial \alpha},\,\,\,\,\pi_{T}=\frac{\partial S}{\partial T},\label{e23}\\
\dot{\alpha}=\frac{\partial H}{\partial \pi_{\alpha} },\,\,\,\,\dot{T}=\frac{\partial H}{\partial \pi_{T} },\,\,\,\,\frac{\partial H}{\partial t}=-\frac{\partial \mathcal{L}}{\partial t}\label{e23-1}
\end{eqnarray}
\end{subequations}
\begin{equation}\label{e24}
H(\alpha,T;\pi_\alpha,\pi_T;t)+\frac{\partial S(\alpha, T, t)}{\partial t}=0,
\end{equation}
 the generalized momenta are written by,
\begin{eqnarray}\label{e25}
	\pi_{\alpha} = \frac{\partial \mathcal{L}}{\partial\dot{\alpha}}= -6\dot{\alpha}e^{3\alpha},\,\,\,\,\,
	\pi_T = \frac{\partial \mathcal{L}}{\partial\dot{T}}= \dot{T}e^{3\alpha}.
\end{eqnarray}
On the other hand, the canonical Hamiltonian will obtain by aforesaid equations as,
\begin{equation}\label{e26}
H=\pi_\alpha \dot{\alpha}+\pi_T \dot{T}-\mathcal{L}=\left(-\frac{\pi_{\alpha}^2}{12}+ \frac{\pi_T^2}{2}+e^{6 \alpha} \widetilde{V}-3 k e^{4 \alpha}\right)e^{-3 \alpha}.
\end{equation}
Now using Eqs. \eqref{e24}, \eqref{e25} and \eqref{e26} we obtain the following Hamilton equation
\begin{equation}\label{e27}
-\frac{1}{12}\left(\frac{\partial S}{\partial \alpha}\right)^2+\frac{1}{2}\left(\frac{\partial S}{\partial T}\right)^2+e^{6\alpha}\widetilde{V}(T)-3k e^{4\alpha}=0.
\end{equation}
As we know, the choice of closed string tachyon potential plays the role of an important in String Theory. But we are going to perform the current model for an cosmological analysis. Therefore, different suggestions expressed for selecting of the corresponding potential \cite{ADAC, MRGA}. In that case, we will extend the job \cite{EscamillaRivera:2011di} by taking the function $S$ in a non-flat universe by curvature $k$ in the following form,
\begin{equation}\label{e28}
S(\alpha, T, t)= e^{3\alpha(t)} \left[W(T(t))+\beta k Z(T(t))\right],
\end{equation}
where $W(T)$ and $Z(T)$ are an arbitrary function with dependence on $T$, and $\beta$ is a constant coefficient. Now by substituting \eqref{e28} into \eqref{e27}  the effective tachyon potential is written in terms of functions $W(T)$ and $Z(T)$ as,
\begin{eqnarray}\label{e29}
\begin{aligned}
\widetilde{V}(T)&=&\frac{3}{4}\left(W+\beta k Z \right)^2-\frac{1}{2}\left(\partial_T W+\beta k\, \partial_T Z\right)^2+3 k e^{-2\alpha}.
\end{aligned}
\end{eqnarray}

%##########################################################################################

\section{Reconstructing Closed String Tachyon Potential}\label{s4}

In this section we are going to describe the cosmological evolution of our model with the closed string tachyon coupled to  a modified Chaplygin gas. Let us remark that the recent superstring corrections interpreted compactification on internal manifold that internal curvature is everywhere negative \cite{GSH, MRDR}. Therefore, from the point of view $4 d$ geometry, the internal curvature is a negative constant $\mathcal{R}<0$, i.e., the internal curvature is not a functional of $T$.\\

In order to obtain effective tachyon potential, we take the function $W(T)$ in the form,
\begin{equation}\label{e30}
W(T)=C+D T^2,
\end{equation}
 where $C$ and $D$ are constant coefficients in which they play a role of important for description cosmological solution. The motivation of this choice is based on $a)$ crossing of Equation of State (EoS) over phantom-divide-line, and $b)$ achieve to a polynomial function for tachyonic potential as mentioned in Ref. \cite{Yang:2005rw, EscamillaRivera:2011di}. \\
 In this model, we simplicity take $W=Z$, then by inserting \eqref{e30} into \eqref{e29}, the effective tachyon potential is yielded as,
 \begin{eqnarray}\label{e31}
\widetilde{V}(T)&=&\frac{3}{4}D^2\left(1+ \beta k\right)^2 T^4+D \left(\frac{3}{2}C-2D\right) \left(1+ \beta k\right)^2 T^2
+\left(\frac{3}{4}C^2 \left(1+ \beta k\right)^2+3 k e^{-2\alpha}\right).
\end{eqnarray}
As we know, the constant curvature of internal manifold $\mathcal{R}$ is not a functional of tachyon field. Now with correspondence of Eqs. \eqref{e31} and \eqref{e6}, the negative curvature is given by,
\begin{equation}\label{e32}
\mathcal{R}=-\frac{3}{2}C^2 \left(1+ \beta k\right)^2,
\end{equation}
therefore, the closed string tachyon potential is reduced to,
\begin{equation}\label{e33}
V(T)=\frac{3}{4}D^2\left(1+ \beta k\right)^2 T^4+D \left(\frac{3}{2}C-2D\right) \left(1+ \beta k\right)^2 T^2+3 k e^{-2\alpha}.
\end{equation}
Making use of the momentum related to $\alpha$ and
$T$ Eqs. \eqref{e23} and the ansatz for $S$ Eq. \eqref{e28} and $W$ Eq. \eqref{e30}, we can obtain the tachyon field solution and the
scale factor function in terms of time:
\begin{equation}\label{e34}
T=e^{2D(1+\beta k)t},
\end{equation}
\begin{equation}\label{e35}
a(t)=exp \left[{-\frac{1}{2}C(1+\beta k)t-\frac{1}{8}e^{4D(1+\beta k)t}}\right].
\end{equation}
We note that Eqs. \eqref{e34} and \eqref{e35} are strictly constrained to the values of $C$ and $D$, which ones will play an important role for the description of the cosmological evolution. In next section, we intent to investigate the effect of obtained parameters with Chaplygin gas.

%##########################################################################################

\section{Interacting closed string tachyon with Chaplygin gas}\label{s5}
In this section, we consider an interaction between the closed string tachyon and Chaplygin gas. In connection with string theory, the equation of state of the Chaplygin gas has obtained from the Nambu-Goto action for a D-brane moving in a $(D+2)$-dimensional space-time in the light cone parametrization \cite{HBBE, RJAK, NOGA}. The equation of state the modified Chaplygin gas is given by,
\begin{equation}\label{e9}
p_{MCG}=A \rho_{MCG}-\frac{B}{\rho_{MCG}^{\gamma}},
\end{equation}
where $p_{MCG}$ and $\rho_{MCG}$ are the pressure and energy density of modified Chaplygin gas where $A$ and $B$
are positive constants and $0 \leq \gamma \leq 1$. Therefore, the total energy density and pressure are given respectively by,
\begin{equation}\label{e10}
\rho_{tot}=\rho_T+\rho_{MCG},
\end{equation}
\begin{equation}\label{e11}
p_{tot}=p_T+p_{MCG}.
\end{equation}
As we know the continuity equation derived from $T^{\mu \nu}\,_{;\nu} = 0$, then the general form of continuity equation is,
\begin{equation}\label{e11-1}
    \dot{\rho}_{tot}+3 H (\rho_{tot}+p_{tot})=0,
\end{equation}
now, by taking an energy flow between closed string tachyon and Chaplygin gas, we have to introduce a phenomenological coupling function in terms of product of the Hubble parameter and the energy density of the Chaplygin gas. In that case, continuity equations of the closed string tachyon and Chaplygin gas are written respectively by,
\begin{equation}\label{e12}
\dot{\rho}_{T}+3 H (\rho_T+p_T)=-Q,\\
\end{equation}
\begin{equation}\label{e13}
\dot{\rho}_{MCG}+3 H (\rho_{MCG}+p_{MCG})=Q,
\end{equation}
where the quantity $Q$ is the interaction term between tachyon field and the Chaplygin gas and one is equivalent to $Q = 3 b^2 H \rho_{MCG}$, where $b^2$ is the coupling parameter or transfer strength \cite{GZKZ}. We note that the interaction term $Q$ has widely described in the literature \cite{CEJ, CLP, NSO, WHC}. This choice is based on positive motivation $Q$, because from the observational data
at the four years WMAP implies that the coupling parameter must be a small positive value \cite{FCal, DNSP}.\\

 By substituting \eqref{e9} into \eqref{e13} we can obtain energy density of modified Chaplygin gas as the following form,
 \begin{eqnarray}\label{e14}
\rho_{MCG}=\left[\frac{B}{\eta}+c_0 \,a^{-3\eta(\gamma+1)}\right]^{\frac{1}{\gamma+1}}.
\end{eqnarray}
where $c_0$ is a constant integral, and employing this expression in Eq. \eqref{e9} we can rewrite the pressure for the Chaplygin gas as:
 \begin{eqnarray}\label{e15}
p_{MCG}=A\left[\frac{B}{\eta}+a^{-3\eta(\gamma+1)}\right]^{\frac{1}{\gamma+1}}-
\frac{B}{\left[\frac{B}{\eta}+a^{-3\eta(\gamma+1)}\right]^{\frac{\gamma}{\gamma+1}}},
\end{eqnarray}
where $\eta=1-b^2+A$. If $\eta << 1$ the universe
undergo a collapse stage at late times, where Eq. \eqref{e15} is negative and Eq. \eqref{e14} decreases in the expansion \cite{EscamillaRivera:2011di}.\\
\begin{figure}[t]
\begin{center}
\subfigure
{\includegraphics[scale=.35]{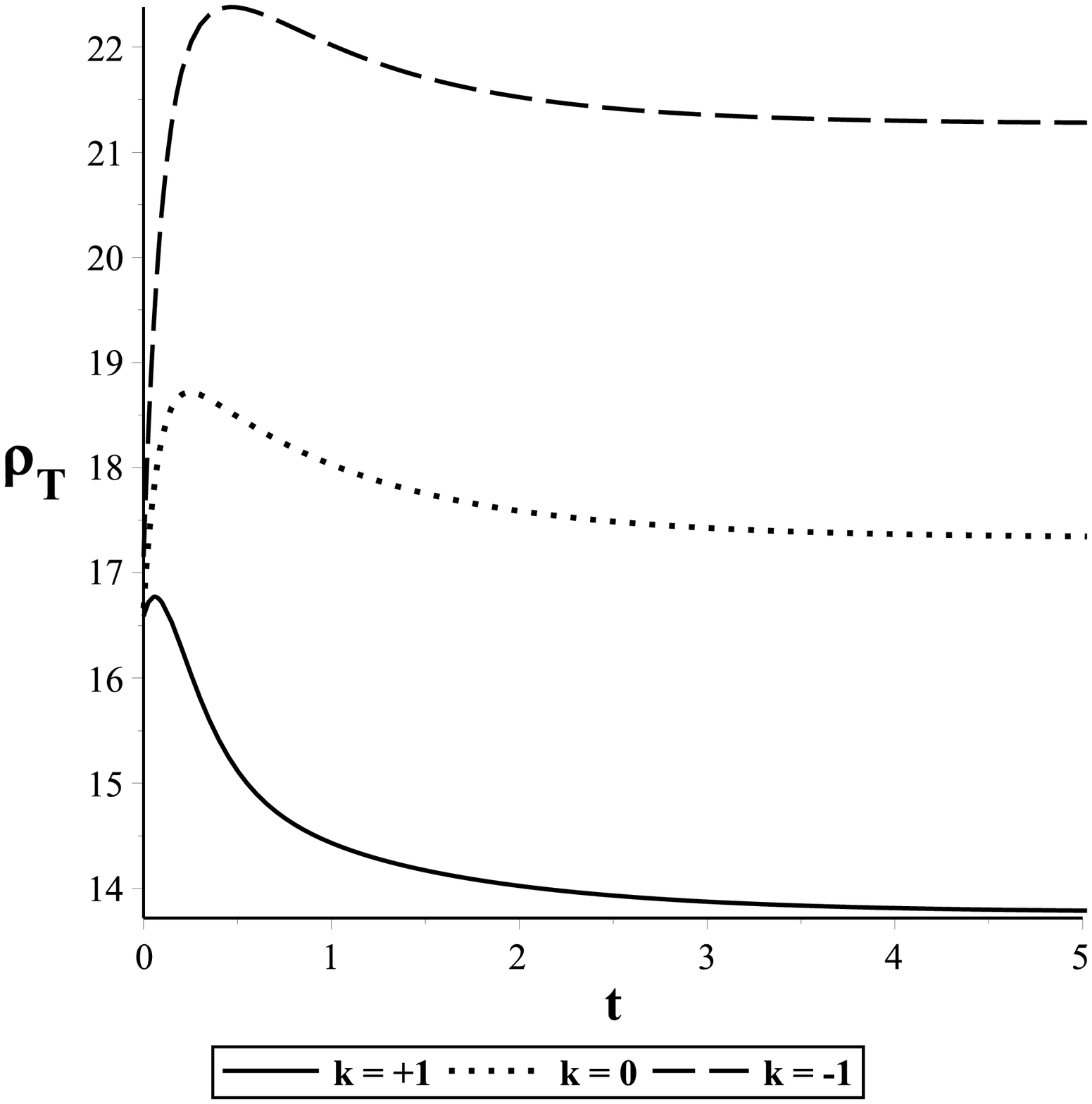}\label{fig1-1}}
\subfigure
{\includegraphics[scale=.35]{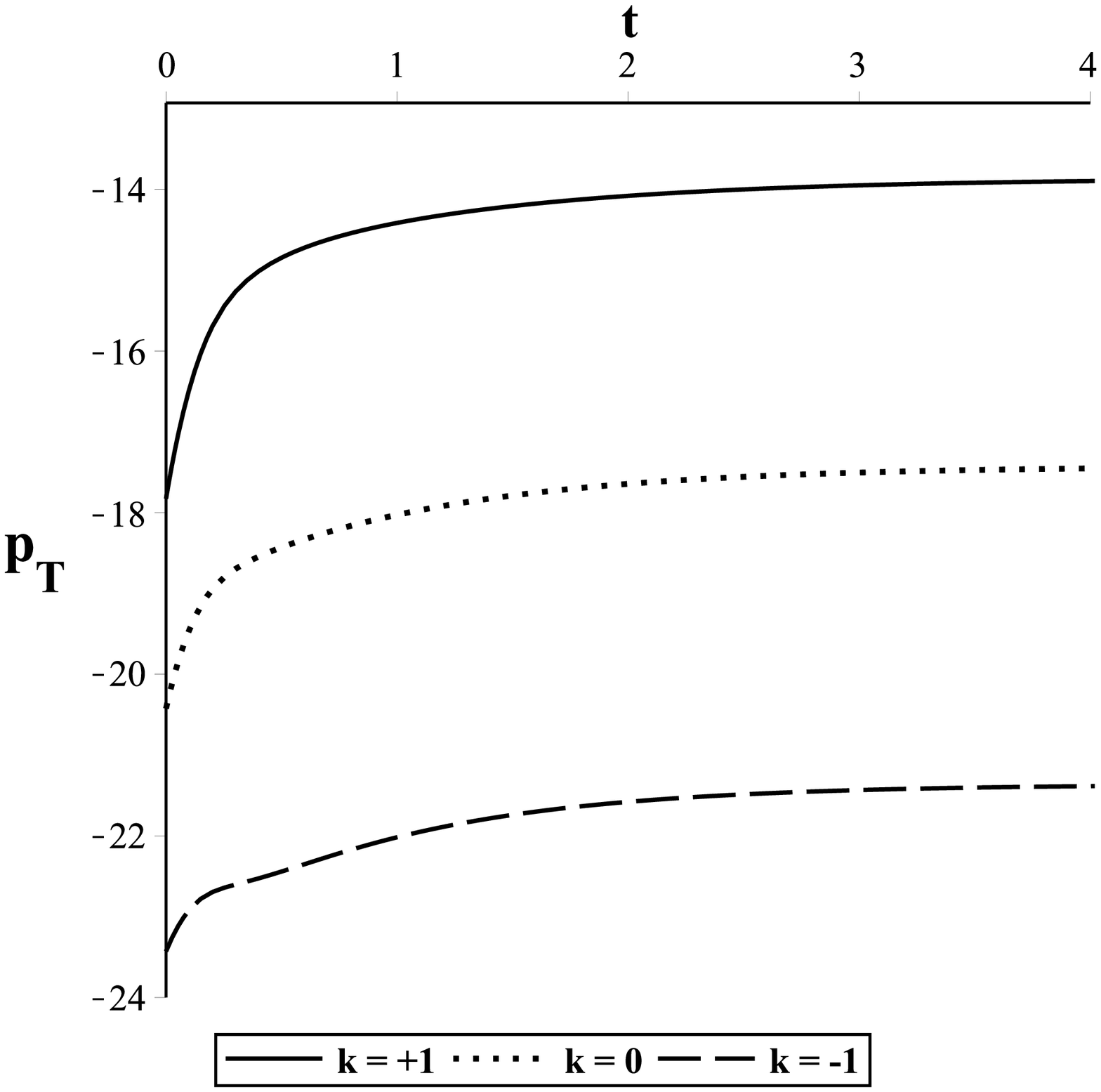}\label{fig1-2}}
\caption{The energy density and pressure of closed string tachyon in terms of time evolution for $B = 2, C = -5, D = -0.25, b = 0.25, \beta = -0.1, \gamma = 0.5, c_0 = 3.25$ and $A = 0.25$ in different cases $k=\pm 1,0$.}\label{fig1}
\end{center}
\end{figure}
Finally, considering the previous background, the expressions for the energy density and pressure of closed string tachyon field coupled to Chaplygin gas are
\begin{eqnarray}\label{e19}
\rho_T&=&3\left(\dot{\alpha}^2+\frac{k}{e^{2\alpha}}\right)-\left[\frac{B}{\eta}+a^{-3\eta(\gamma+1)}\right]^{\frac{1}{\gamma+1}},
\end{eqnarray}
\begin{eqnarray}\label{e20}
p_{T}&=&-\left(2\ddot{\alpha}+3\dot{\alpha}^{2}+\frac{k}{ e^{2\alpha}}\right)-
A\left[\frac{B}{\eta}+a^{-3\eta(\gamma+1)}\right]^{\frac{1}{\gamma+1}}+
\frac{B}{\left[\frac{B}{\eta}+a^{-3\eta(\gamma+1)}\right]^{\frac{\gamma}{\gamma+1}}}.
\end{eqnarray}
By reinserting \eqref{e35} into Eqs. \eqref{e19} and \eqref{e20} the EoS of the closed string tachyon is obtained as,
\begin{equation}\label{e36}
\omega_T=\frac{p_T}{\rho_T}=\frac{-\left(2\ddot{\alpha}+3\dot{\alpha}^{2}+\frac{k}{ e^{2\alpha}}\right)-
A\,\rho_{MCG}+
\frac{B}{\rho_{MCG}^{\gamma}}}{3(\dot{\alpha}^2+\frac{k}
{e^{2\alpha}})-\rho_{MCG}}.\\
\end{equation}
\begin{figure}[t]
\begin{center}
\includegraphics[scale=.4]{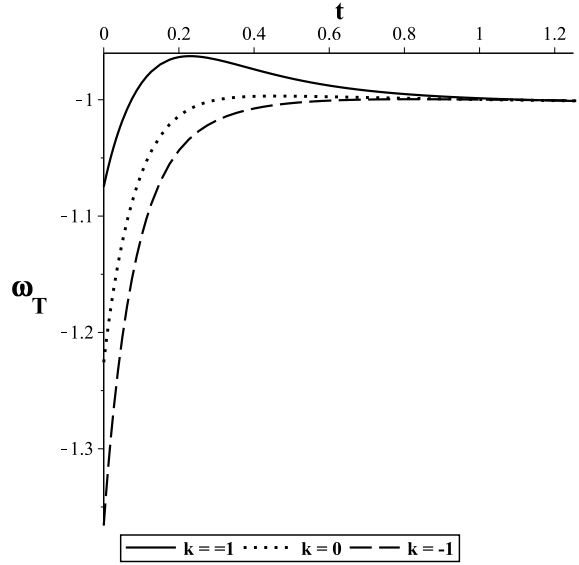}
\caption{The EoS of closed string tachyon in terms of time evolution for $B = 2, C = -5, D = -0.25, b = 0.25, \beta = -0.1, \gamma = 0.5, c_0 = 3.25$ and $A = 0.25$ in different cases $k=\pm 1,0$.}\label{fig2}
\end{center}
\end{figure}
\begin{figure}[h]
\begin{center}
\subfigure
{\includegraphics[scale=.27]{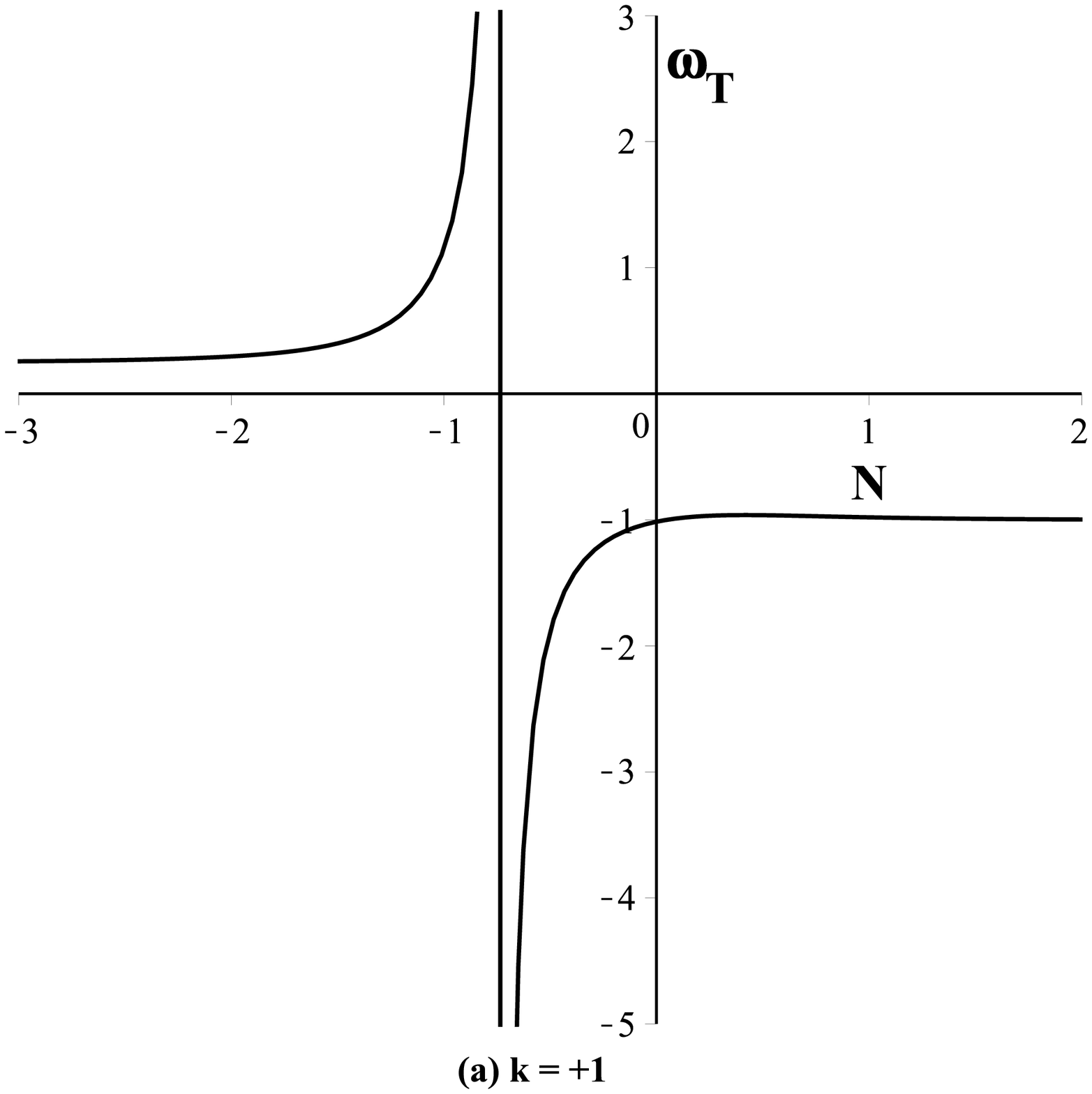}\label{fig3-1}}
\subfigure
{\includegraphics[scale=.27]{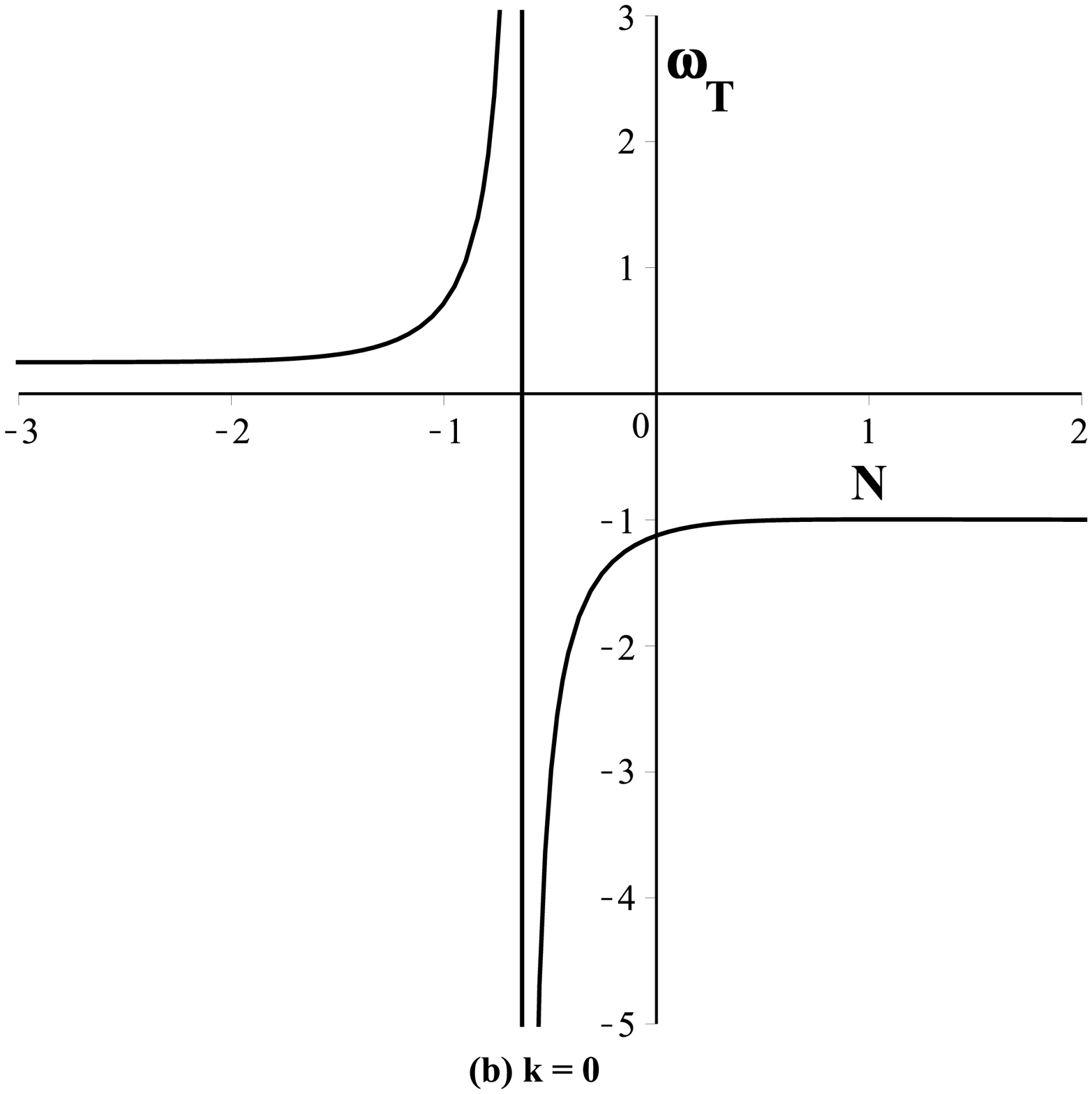}\label{fig3-2}}
\subfigure
{\includegraphics[scale=.27]{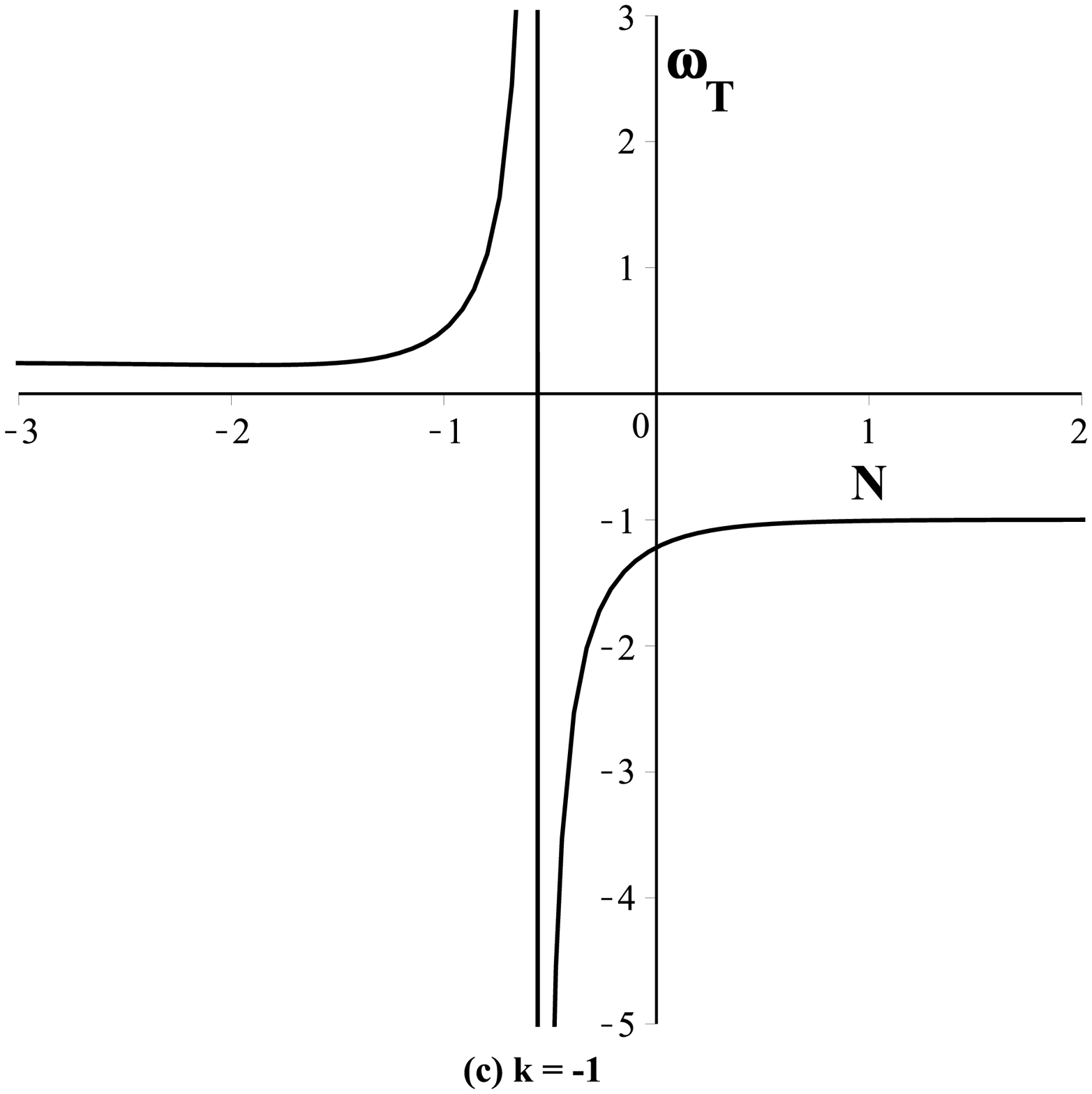}\label{fig3-3}}
\caption{The EoS of closed string tachyon in terms of e-folding number for $B = 2, C = -5, D = -0.25, b = 0.25, \beta = -0.1, \gamma = 0.5, c_0 = 3.25$ and $A = 0.25$ in different cases $k=\pm 1,0$.}\label{fig3}
\end{center}
\end{figure}
We can see variation of the cosmological parameters against time evolution by interacting Chaplygin gas with the closed string tachyon by geometries ($k=0,\pm1$) in the Figures \ref{fig1} and \ref{fig2}. We note that chosen coefficients play the role of an important to plot the cosmological parameters such as the energy density and pressure. Then, the motivation of the selections is based on crossing EoS over phantom-divide-line, positivity energy density and negativity pressure.\\

Since in this paper we use the natural units as $c = \hbar = m_p = 1$, therefore in order to have a more complete discussion, we represent the free parameters of the model in terms of observable quantities. For this purpose, to have an accelerated expansion we draw the EoS of the closed string tachyon versus the e-folding number, $N = ln(a)$ as the time variable in Fig. \ref{fig3}. We can see the values of EoS of the closed string tachyon in three cases $N \rightarrow -\infty$, late time ($N = 0$) and $N \rightarrow +\infty$ respectively with values $0.25$, $-1.017$ and $-1.006$ for geometry $k = +1$ in Fig. \ref{fig3-1}, and $0.25$, $-1.125$ and $-1.005$ for geometry $k = 0$ in Fig. \ref{fig3-2}, and $0.25$, $-1.221$ and $-1.004$ for geometry $k = -1$ in Fig. \ref{fig3-3}. we noted that when universe is undergoing an accelerated expansion, the EoS of the closed string tachyon crosses the value of $-1$ in late time where in the scenario it even can be seen for different geometries ($k = 0,\pm1$) in Fig. \ref{fig3}.

%#######################################################################################3

\section{Conditions for an Accelerated Universe and the Stability Analysis}\label{s6}

In this section, we are going to investigate two issues: first, the condition of an accelerated universe in our model and second, the stability analysis of aforesaid proposal. \\

On one hand, we study on the condition of expanding universe accelerating for the closed string tachyon.
For this purpose, we can obtain $\rho_T$, and $p_T$ by inserting \eqref{e35} into Eqs. \eqref{e19} and \eqref{e20} as follows,
\begin{eqnarray}\label{e37}
\rho_T&=&\frac{3}{4}(1+\beta\,k)^2\,\left[C+D\,e^{4D(1+\beta k)t}\right]^2 +3k \exp{\left[C(1+\beta k)t+\frac{1}{4}e^{4D(1+\beta k)t}\right]} \nonumber \\
 &&-\left(\frac{B}{\eta}+\exp\left[\frac{3}{2}C \eta (\gamma+1)(1+\beta k)t+\frac{3}{8}\eta(\gamma+1)e^{4D(1+\beta k)t}\right]\right)^{\frac{1}{\gamma+1}},
\end{eqnarray}
\begin{eqnarray}\label{e38}
&p_{T}=(1+\beta\,k)^2\Big[4 D^2 e^{4 D(1+\beta k)t}-\frac{3}{4} \left(C+D e^{4D(1+\beta k)t}\right)^2 \Big]
 -k \exp{\left[C(1+\beta k)t+\frac{1}{4}e^{4D(1+\beta k)t}\right]}\nonumber \\
&-A\left(\frac{B}{\eta}+\exp\left[\frac{3}{2}C \eta (\gamma+1)(1+\beta k)t+\frac{3}{8}\eta(\gamma+1)e^{4D(1+\beta k)t}\right]\right)^{\frac{1}{\gamma+1}}\nonumber\\
&+\frac{B}{\left[\frac{B}{\eta}+\exp\left[\frac{3}{2}C \eta (\gamma+1)(1+\beta k)t+\frac{3}{8}\eta(\gamma+1)e^{4D(1+\beta k)t}\right]\right]^{\frac{\gamma}{\gamma+1}}}.
\end{eqnarray}
Now we can find a constraint for the accelerated universe by weak energy condition ($\rho >0$ and $p_{T}+\rho_T>0$) in terms of current epoch time $t_0$, in the following form,
\begin{eqnarray}\label{e39}
&&\frac{3}{4}(1+\beta\,k)^2\,\left(C+D\,e^{4D(1+\beta k)t_0}\right)^2+3k \exp{\left[C(1+\beta k)t_0+\frac{1}{4}e^{4D(1+\beta k)t_0}\right]} \nonumber \\
&&-\left(\frac{B}{\eta}+\exp\left[\frac{3}{2}C \eta (\gamma+1)(1+\beta k)t_0+\frac{3}{8}\eta(\gamma+1)e^{4D(1+\beta k)t_0}\right]\right)^{\frac{1}{\gamma+1}} > 0,
\end{eqnarray}
and
 \begin{eqnarray}\label{e40}
 \begin{aligned}
&4 D^2 (1+\beta\,k)^2\,e^{4 D(1+\beta k)t_0}+2 k \exp{\left(C(1+\beta k)t_0+\frac{1}{4}e^{4D(1+\beta k)t_0}\right)}\\
&-(A+1)\left[\frac{B}{\eta}+\exp\left(\frac{3}{2}C \eta (\gamma+1)(1+\beta k)t_0+\frac{3}{8}\eta(\gamma+1)e^{4D(1+\beta k)t_0}\right)\right]^{\frac{1}{\gamma+1}}\\
&+\frac{B}{\left[\frac{B}{\eta}+\exp\left[\frac{3}{2}C \eta (\gamma+1)(1+\beta k)t_0+\frac{3}{8}\eta(\gamma+1)e^{4D(1+\beta k)t_0}\right]\right]^{\frac{\gamma}{\gamma+1}}} > 0,
\end{aligned}
\end{eqnarray}
Eqs. \eqref{e39} and \eqref{e40} are a constraints for all the coefficients of the model in the current epoch of the universe.\\
\begin{figure}[t]
\begin{center}
\subfigure
{\includegraphics[scale=.27]{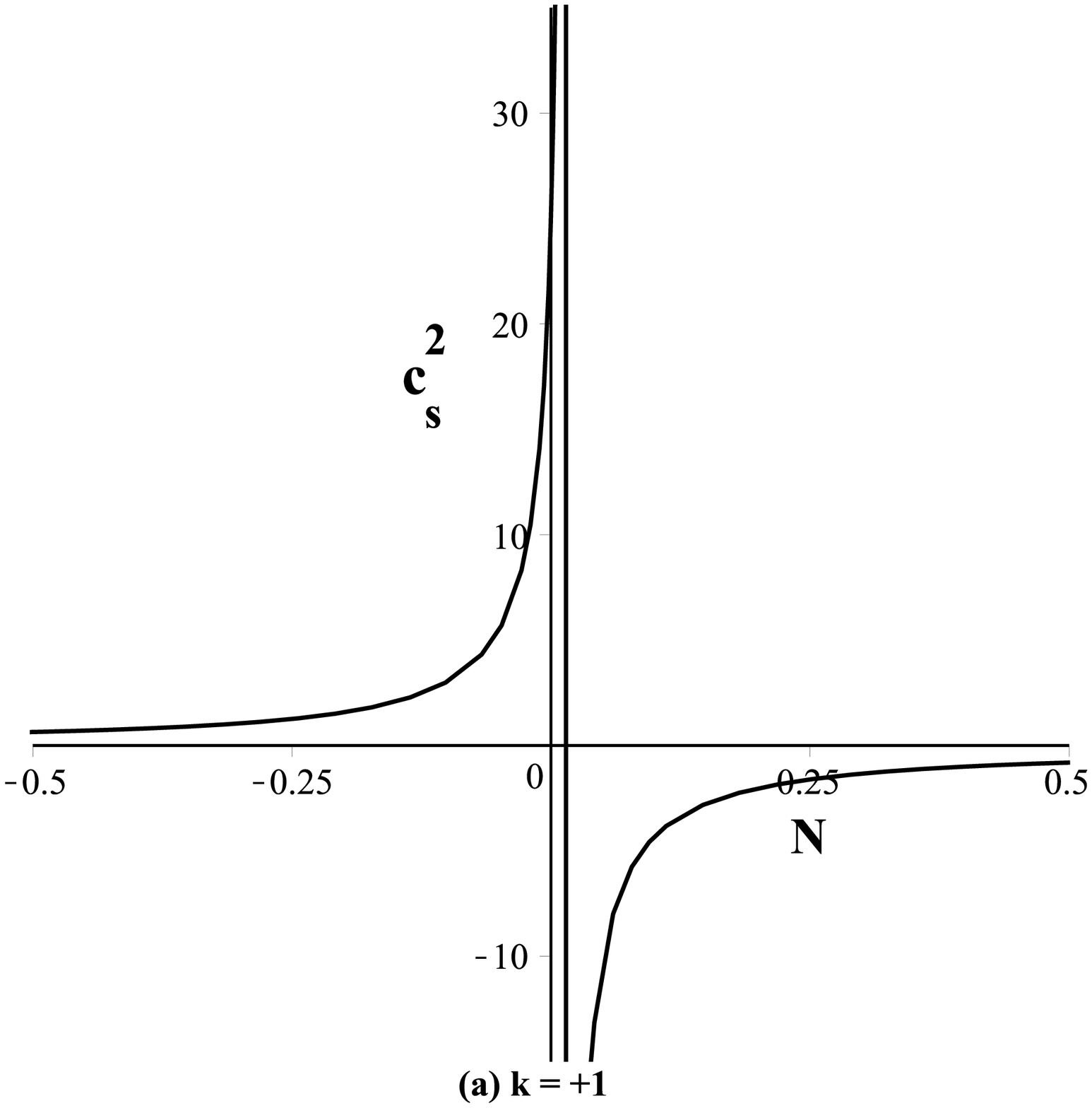}\label{fig4-1}}
\subfigure
{\includegraphics[scale=.27]{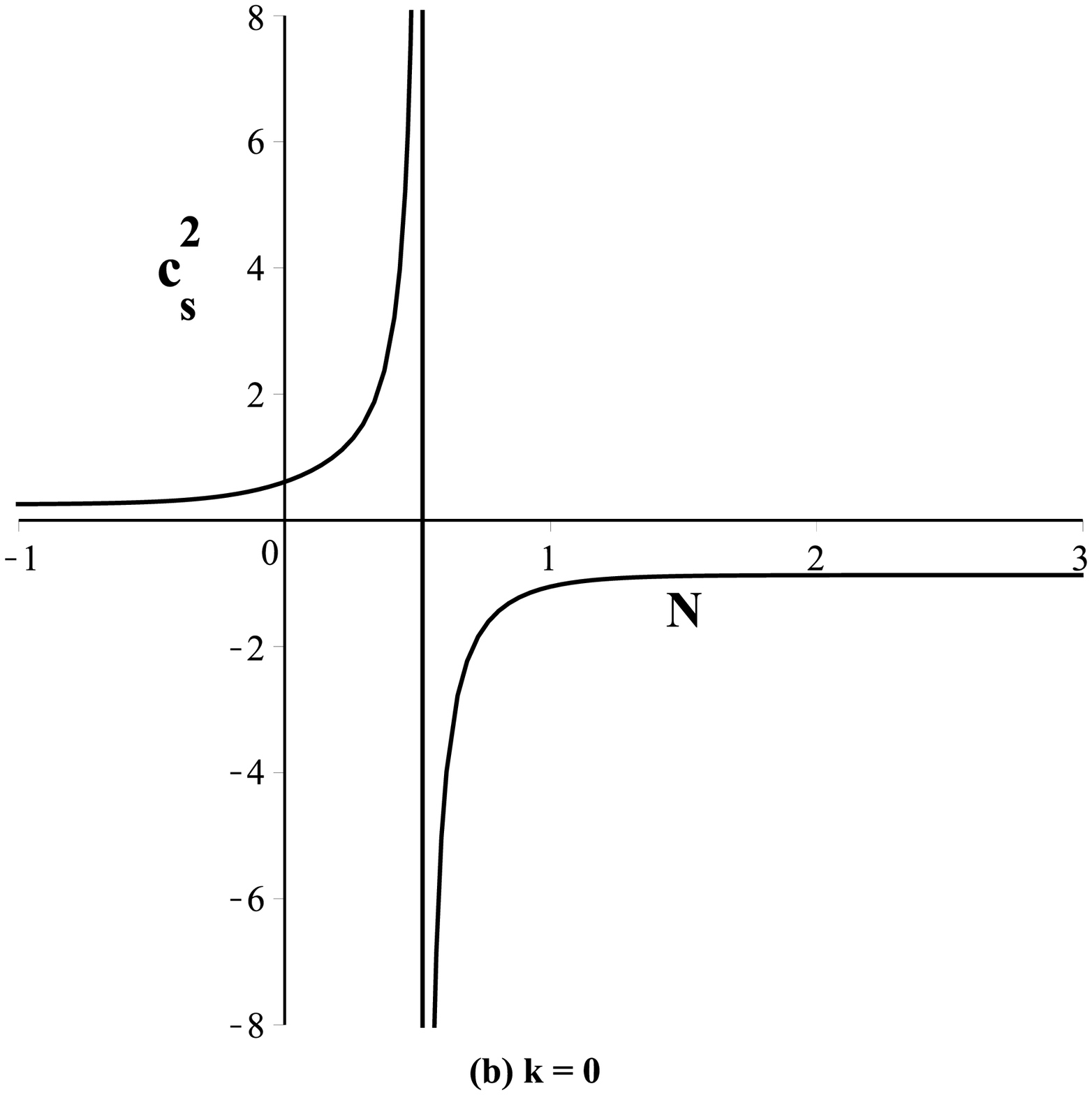}\label{fig4-2}}
\subfigure
{\includegraphics[scale=.27]{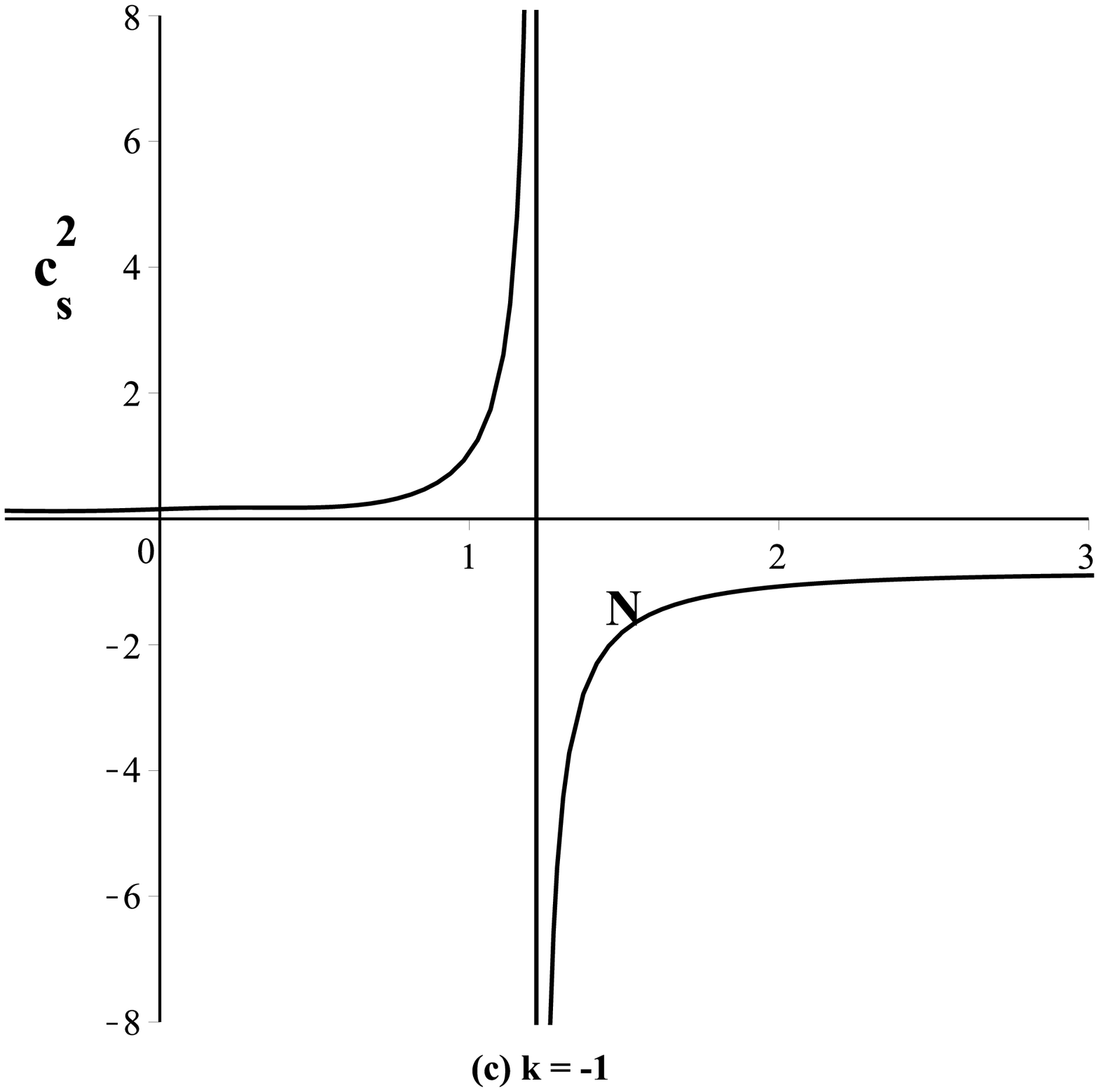}\label{fig4-3}}
\caption{Graphs of the $c_s^2$ in terms of time evolution for $B = 2, C = -5, D = -0.25, b = 0.25, \beta = -0.1, \gamma = 0.5, c_0 = 3.25$ and $A = 0.25$ in geometries $k=\pm1, 0$.}\label{fig4}
\end{center}
\end{figure}
On the other hand, we will discuss the stability of our model with presence of closed string tachyon field.
In that case, we will describe the corresponding stability with an useful function $c^2_s=\frac{\partial p_T}{\partial \rho_T} = \frac{\dot{p}_T}{\dot{\rho}_T}$. The stability condition occurs when the function $c^2_s$ becomes bigger than zero. Of course this function represent sound speed in a perfect fluid.\\

 We noted that a general thermodynamic system can be described with adiabatic and non-adiabatic perturbations by three variables, $\rho_T$, $p_T$ and $S$ (entropy). If we consider $p_T = p_T(S,\rho_T)$, so the pressure perturbation can be written as: $\delta p_T = \frac{\partial p_T}{\partial S} \delta S+\frac{\partial p_T}{\partial \rho_T} \delta \rho_T =  \frac{\partial p_T}{\partial S} \delta S+c^2_s\, \delta \rho_T$. The first term be related to non-adiabatic system, and $c^2_s$ in the second term be related to adiabatic sound speed, i.e., system is when adiabatic in which $\delta S = 0$. Therefore, we describe the stability of our model just by adiabatic sound speed.\\

In this way, by making derivative Eqs. \eqref{e19} and \eqref{e20} with respect to time evolution and numerical computing the function $c^2_s$ in terms of time evolution, we can plot speed sound function by various geometries ($k=0,\pm1$) in Figure \ref{fig4}.\\

We can see the values of $c^2_s$ in three cases $N \rightarrow -\infty$, late time ($N = 0$) and $N \rightarrow +\infty$ respectively with values $0.25$, $25.25$ and $-0.867$ for geometry $k = +1$ in Fig. \ref{fig4-1}, and $0.25$, $0.609$ and $-0.867$ for geometry $k = 0$ in Fig. \ref{fig4-2}, and $0.25$, $0.154$ and $-0.867$ for geometry $k = -1$ in Fig. \ref{fig4-3}.\\

 Therefore, the Fig. \ref{fig4} shows us that there is stability in late time for every three universe of closed, open and flat, because values $c^2_s$ are positive for every three universe in late time (i.e., $N=0$).
%####################################################################################

\section{Conclusions}\label{s7}

In this paper, we have studied closed string tachyon with a constant dilaton field in $26 d$ space-time for describing something mysterious in the cosmology. To understand this issue, we have considered the corresponding model by interacting with modified Chaplygin gas. We noted that the corresponding action has been written as an effective four-dimensional action by compactification on a non-flat internal $22$ d, in which the internal compact space considered a constant volume. The Einstein and field equations have been obtained and by taking an interaction between the closed string tachyon with modified Chaplygin gas we could find the energy density and pressure of closed string tachyon.\\

By using canonical Hamiltonian analysis and the corresponding action, we obtained the continuity equation and then the effective tachyon potential have been found in terms of an arbitrary function ($W(T)$ and $Z(T)$) proportional to tachyon field. In order to reconstruct closed string tachyon potential, we took arbitrary function $W(T)$ such as a quadratic function of tachyon field. In additional, by employing canonical Hamiltonian equations we obtained the tachyon field and the scale factor in terms of time evolution. One of cosmology characteristics that confirm observational data is based on crossing the EoS from phantom-divide-line, in which one calculated in terms of time evolution and e-folding number. Also we plotted the EoS with respect to time evolution and e-folding number for various geometries. The graph of EoS showed accelerating universe and cross over phantom-divided line. Next we obtained a constraint by weak energy condition. Finally we have considered stability analysis for the presented model by using an useful function called the sound speed. This function is employed in a perfect fluid, in which its value is greater than zero. We plotted variation of the sound speed versus time evolution and the corresponding graphs showed stability in late time. The interesting problem here was to consider the model with the curvature of the internal space in a non-constant internal volume scenario.

%#########################################################################################

\section{Acknowledgements}
C. Escamilla-Rivera is supported by Fundaci\'on Pablo Garc\'ia and FUNDEC, M\'exico.\\
Also the authors would like to thank an anonymous referee for crucial remarks and advices.

%#########################################################################################

%##################################################################################

\end{document}